\documentstyle[twocolumn,amssymb,eqsecnum,aps]{revtex}

\begin{document}
\draft
\title{Ohm's law revision}
\author{M. V. Cheremisin$^{*}$}
\address{A.F.Ioffe Physical-Technical Institute,St.Petersburg, Russia}
\date{\today}
\maketitle

\begin{abstract}
The standard ohmic measurements by means of two extra leads contain an
additional thermal correction to resistance. The current results in
heating(cooling) at first(second) sample contact due to Peltier effect. The
contacts temperatures are different. The measured voltage is the sum of the
ohmic voltage swing and Peltier effect induced thermopower which is $linear$
on current. As a result, the thermal correction to resistance measured
exists at $I\rightarrow 0$. The correction should be in comparison with
ohmic resistance. Above some critical frequency dependent on thermal
inertial effects the thermal correction disappears. 
\end{abstract}


\preprint{HEP/123-qed}\narrowtext

As well known the ohmic measurements (Fig.\ref{Fig.1}) are carrying out at
low current density in order to prevent heating. Usually, the only Joule
heat is considered to be important. We recall, in contrast to Joule heat,
the Peltier and Thomson effects are linear on current. The crucial point of
the present paper is that the linear on current Peltier effect influences
the ohmic measurements and results in the correction to resistance measured.
It's well known that the one of the sample contacts is heated while the
another one is cooled due to current induced Peltier effect. The temperature
gradient established is proportional to current. In this case the Thomson
heat is proportional to second power on current, hence, could be neglected.
Finally, the voltage swing across the circuit includes Peltier effect
induced thermopower which is linear on current. Accordingly, there exists
the thermal correction to ohmic resistance of the sample.

Let us first consider an isotropic(or of cubic symmetry) conductor which can
be in the thermodynamic non-equilibrium with respect to conducting
electrons. In general, for inhomogeneous conductor the current density, $%
{\bf j}$ ,\hspace{0in}and energy flux density, ${\bf q}$, are given by the
following equations$^{\cite{Landau3}}$. 
\begin{equation}
{\bf j}=\sigma ({\bf E-}\alpha {\bf \nabla }T),  \eqnum{1}  \label{current}
\end{equation}
\begin{equation}
{\bf q}=(\phi {\bf +}\alpha T){\bf j}-\varkappa {\bf \nabla }T.  \eqnum{2}
\label{heat}
\end{equation}
Here, $\sigma $ is the conductivity, $\alpha $ is the thermopower, $%
\varkappa $ is the thermal conductivity. For an inhomogeneous conductor the
electrochemical potential, $\phi =\varphi +\mu /e$ , is the sum of the
electric potential, $\varphi $, and chemical potential, $\mu $, of the
conducting electrons. Note, for homogeneous conductor the above definition
of the potential results in an additional unimportant constant to $\varphi $%
, therefore, the average microscopic electric field, $-{\bf \nabla }\varphi $%
, coincides with ${\bf E=}-{\bf \nabla }\phi $. The first term in Eq.(\ref
{current}) corresponds to conventional Ohm's law, the second one describes
the thermoelectric phenomena. For steady state 
\begin{equation}
\text{div}{\bf j}=0,  \eqnum{3}  \label{SS_current}
\end{equation}
\begin{equation}
Q=-\text{div}{\bf q}=\text{div(}\varkappa {\bf \nabla }T)+j^{2}/\sigma
-jT\nabla \alpha =0,  \eqnum{4}  \label{SS_heat}
\end{equation}
where $Q$ is the total amount of heat, evolved per unit time and volume of
the conductor. The current flow is accompanied by both the Joule and Tomson
heats which are proportional to the second(first) power of current
respectively. Using Eqs.(\ref{current}-\ref{SS_heat}) one should find the
potential, $\phi ({\bf r})$ , and temperature, $T({\bf r})$ , for the
conductor under given boundary conditions.

We now consider the thermal effects in connection with ohmic measurements
(see Fig.(\ref{Fig.1})) of a conductor resistance. The current carrying
conductor is connected by means of two identical extra leads to the current
source (not shown). We assume that both contacts are ohmic ones. Then, $%
\alpha $, $\sigma $, $\varkappa $, the length, $l$, and the conductor
cross-section, $S$, are different for the leads and the sample. The voltage
is measured between the open ends (''c'' and ''d'') being kept at
temperature, $T_{0}$ , of an external thermal reservoir. In general, the
contacts (''a'' and ''b'') could be at different temperatures $T_{a}$ and $%
T_{b}$ respectively.

As well known the Peltier heat is generated by the current crossing the
contact of two different conductors. At the contact( let say ''a'' in Fig.(%
\ref{Fig.1})) the temperature $T$ , the electrochemical potential $\phi $,
the normal components of the current $I=jS$ and the total energy flux $qS$ ,
are all continuous. Then, there exists the difference, $\Delta \alpha
=\alpha _{1}-\alpha _{2}$, of thermopowers. For $\Delta \alpha >0$ the
charge intersecting the contact ''a'' gains the energy $e\Delta \alpha T_{a}$%
. Consequently, $Q_{a}=I\Delta \alpha T_{a}$ is the amount of Peltier heat
evolved per unit time in the contact ''a''. We underline that $Q_{a}$ could
be calculated by another way accounting the Thomson term in Eq.(\ref{SS_heat}%
). Indeed, $Q_{a}\equiv \int -IT\nabla \alpha dx$, where the integration is
accomplished over the contact length. In fact, Peltier effect is equivalent
of Thomson effect established at the contact.

For $\Delta \alpha >0$ and current direction shown in Fig.(\ref{Fig.1}) the
contact ''a'' is heated while the contact ''b'' is cooled. Thus, the
contacts are at different temperatures and $T_{a}-T_{b}=\Delta T>0$. We now
show that the standard ohmic measurements always result in the thermal
correction to resistance measured. Using Eq.(\ref{current}), the voltage
swing , $U$ , between the ends ''c'' and ''d'' is 
\begin{equation}
U=\int\limits_{c}^{d}\left( {\bf j/}\sigma +\alpha \bigtriangledown T\right)
dx=RI+\varepsilon _{T},  \eqnum{5}  \label{temf}
\end{equation}
where $R=2R_{1}+R_{2}=2l_{1}/(S_{1}\sigma _{1})+l_{2}/(S_{2}\sigma _{2})$ is
the total resistance of the circuit. The first term in Eq.(\ref{temf})
corresponds to Ohm's law$^{\cite{Ohm}}$. The second term, $\varepsilon
_{T}=\int\limits_{c}^{d}\alpha dT$ , coincides with equation for
conventional thermoelectromotive force under zero current conditions$^{\cite
{Landau3}}$. Note, $\varepsilon _{T}$ is an universal value since for
arbitrary cooling conditions it depends on the contact temperatures only.
It's worth noting that there exists the correlation between
thermoelectromotive force, Peltier and Thomson heats. Indeed, the total
power evolved in the circuit, $UI$ , is the sum of Joule heat $RI^{2}$ and
the thermal effects related power $\varepsilon _{T}I$. Then, the product $%
\varepsilon _{T}I$ is exactly the sum of Peltier heat , $Q_{P}=Q_{a}-Q_{b}=I%
\Delta \alpha \Delta T$ , evolved at both contacts and Tomson heat, $%
Q_{T}=-\int\limits_{c}^{d}IT\nabla \alpha dx$ in the conductors bulk: 
\begin{equation}
I\varepsilon _{T}=Q_{P}+Q_{T}.  \eqnum{6}  \label{equality}
\end{equation}
According to Eq.(\ref{equality}), for arbitrary circuit under the same
contacts temperatures( $T_{a}$, $T_{b}$ and $T_{0}$) the zero current
measurements of thermoelectromotive force allow to find the total amount of
both the Peltier and Thomson heats at $I\neq 0$.

We recall that the sample contacts are always extra heated(cooled) because
of Peltier effect. The difference of contact temperatures $\Delta T$ is
linear on current, thus, there exists the thermal correction to ohmic
resistance $\Delta R=\varepsilon _{T}/I=U/I-R$. For simplicity, we further
assume that the conductivity, $\sigma $ ,thermopower, $\alpha $ , and the
thermal conductivity, $\varkappa $ , are all temperature independent. In
that case, the thermopower is given $\varepsilon _{T}=\Delta \alpha \Delta T$%
.

Using Eqs.(\ref{SS_heat},\ref{temf}) one could easily find out voltage swing 
$U$ and, thus, thermal correction $\Delta R$ for arbitrary circuit. Note,
the real cooling conditions strongly influence $\Delta R$. We now precise
the cooling conditions of the circuit shown in Fig.(\ref{Fig.1}). Let us
consider the adiabatic conditions when the sample is thermally isolated with
respect to environment. As an example, the sample could be placed into the
vacuum chamber(see Fig.\ref{Fig.1}) surrounded by thermal reservoir kept at $%
T_{0}$. Then, we further neglect the heat transfer within the leads.
Actually, that means the sample is self-isolated. We emphasize that under
the above conditions the sample is not heated. Really, under zero on current
approximation $T_{a},T_{b}\approx T_{0}$ , hence, the amount Peltier heat
evolved at the first contact ''a'' is equal to one absorbed at the second
contact ''b''. At both contacts the energy flux $qS$ is continuous, thus

\begin{equation}
Q_{a}=-Q_{b}=I\Delta \alpha T_{0}=-\varkappa _{2}\frac{dT}{dx}S_{2}. 
\eqnum{7}  \label{Condition}
\end{equation}

Using Eq.(\ref{Condition}) one could find out immediately the thermal
correction to resistivity as follows

\begin{equation}
\Delta R=\frac{T_{0}(\Delta \alpha )^{2}l_{2}}{S_{2}\varkappa _{2}}. 
\eqnum{9}  \label{correction0}
\end{equation}
According to Eq.(\ref{temf},\ref{correction0}) $\Delta R$ depends on
reservoir temperature, geometry and heat conductivity of sample. We
underline that the thermal correction is always positive, since the total
amount of Peltier heat $Q_{P}=\Delta RI^{2}>0$.

Let us estimate the magnitude of the thermal correction to resistivity $%
\Delta R$, when both the conductor and leads are metals. At room temperature
the electron heat conductivity and thermopower of electron gas are: $%
\varkappa =L\sigma T$, $\alpha =\frac{\pi ^{2}k}{2e}\xi $, where $L=\frac{%
\pi ^{2}k^{2}}{3e^{2}}$ is the Lorentz number, $\xi =kT/E_{f}<<1$ is
degeneracy parameter. The difference $\Delta \alpha $ is of the order $\frac{%
k\xi }{e}$. Under the above assumptions one should easily find $\Delta
R/R\sim \xi ^{2}<<1$. Thus, thermal correction is small compared with ohmic
resistance because of degeneracy of the electron gas. Note for semimetals(
bismuth, $E_{f}\sim 35$meV ) the thermal correction could be greater. In
contrast to metal, the thermal correction should be $\xi ^{-2}$ times higher
for non-degenerated semiconductor since in that case $\Delta \alpha \approx
\alpha _{2}=\frac{k}{e}(\frac{5}{2}+r-\xi ^{-1})\sim k/e$. Here, $r=3/2$ is
the parameter related to phonon scattering mechanism. Let us consider
non-degenerated n-InSb at T=0.5K. The Fermi energy lies between the
conduction band and shallow donor impurity level $\Delta E_{d}\approx 7$K.
Accordingly, $\Delta \alpha =11\frac{k}{e}$. Then, at low temperatures the
electron heat conductivity is less than phonon related Debye one $\varkappa
_{ph}=0.05\cdot T^{3}$W/cmK.. Finally, for n-InSb with electron
concentration $n=10^{13}$cm$^{-3}$ and mobility $\mu =5\cdot 10^{6}$cm$^{2}/$%
Vs one obtains the thermal correction to resistivity $\Delta R/R\sim 0.01$.

In reality, the cooling conditions could be different from ones assumed. We
now consider more realistic case when the local cooling of the sample is
important. For example, the sample chamber should contain the gas. One could
take into account the cooling effects using Eq.(\ref{SS_heat}) with the
linear term, $-\beta (T-T_{0})$ , included. Here, $\beta $ denotes the
strength of the sample-to-gas thermal exchange. Finally, under zero on
current approximation the temperature downstream the sample(see Fig.(\ref
{Fig.2}, insert)) is given 
\begin{equation}
T(\eta )=\frac{(T_{a}-T_{0})\sinh [\lambda (1-\eta )]+(T_{b}-T_{0})\sinh
[\lambda \eta ]}{\sinh [\lambda ]}+T_{0},  \eqnum{10}  \label{Distribution2}
\end{equation}
where $\eta =x/l_{2}$ is the dimensionless coordinate. According to Eq.(\ref
{Distribution2}), the sample local cooling is governed by the dimensionless
parameter $\lambda =\sqrt{\frac{\beta }{\varkappa _{2}}}l_{2}$. Actually, $%
\lambda $ is the ratio of outgoing and internal(within the sample) heat
fluxes. When $\lambda <<1$ the local cooling could be neglected, hence, $%
T(\eta )$ is linear(Fig.(\ref{Fig.2})). Then, in the opposite case $\lambda
>>1$ of intensive cooling $T(\eta )$ dependence is sharp near the contacts.

The above result allow us to calculate the thermal correction to sample
resistance. Using Eqs.(\ref{temf},\ref{Condition},\ref{Distribution2}) and
omitting the cumbersome algebraic calculations one could calculate the
thermal correction to resistance as follows 
\begin{equation}
\Delta R=\frac{T_{0}(\Delta \alpha )^{2}l_{2}}{S_{2}\varkappa _{2}}\cdot 
\frac{\tanh (\lambda /2)}{(\lambda /2)}.  \eqnum{11}  \label{correction1 }
\end{equation}
For small cooling $\lambda \rightarrow 0$ Eqs.(\ref{correction0},\ref
{correction1 }) coincide. In the opposite case $\lambda \rightarrow \infty $
of strong cooling the difference $\Delta T$ and , thus, thermal correction
decrease(see Fig.\ref{Fig.2}).

We now estimate $\Delta R$ given by Eq.(\ref{correction1 }{\bf ) }for
natural air convection cooling. For sample with typical dimension $d\sim 
\sqrt{S}$ the outgoing thermal flux is given $\varkappa _{gas}\frac{Nu}{S}%
(T-T_{0}),$ where $Nu\sim 10$ is the Nusselt number. For n-InSb sample(0.5$%
\times $0.5$\times $0.5cm) the heat conductivity is $\varkappa _{ph}=0.15$%
W/Kcm( T=293K). Then, assuming the air heat conductivity $\varkappa
_{gas}=2.6$W/cmK one should find $\lambda =14.$ Thus, the thermal correction
to resistivity is approximately 14 times less compared to one in absence of
convection.

We emphasize that both dc and ac ohmic measurements leads to thermal
correction. However, at high frequency $\Delta R$ diminishes due to thermal
inertial effects. In fact, Eq.(\ref{correction0}) are valid below some
critical frequency $f_{cr}=\chi /d^{2}$, where $\chi $ is the temperature
diffusive coefficient of the sample. For example, At T=293K for metal
conductor $\chi =\varkappa /C\simeq 10^{2}$cm$^{2}$/s, where $C$ is the
calorific capacity of the electron gas. Then for typical metal conductor $%
d\sim \sqrt{S}=1$mm, one obtain the critical frequency as $f_{cr}=10^{4}$Hz.
We suggest that the spectral dependence of thermal correction could be used
to estimate the magnitude of the thermal correction.

In conclusion, the ohmic measurements of a conductor resistance contain the
thermal correction caused by Peltier effect. The thermal correction always
exists, while its magnitude depends on actual cooling conditions of the
circuit. Above some critical frequency dependent on thermal inertial effects
the thermal correction disappears.

\acknowledgments
The authors are grateful to M.I.Dyakonov and V.I.Perel for useful
discussions.

\begin{figure}[tbp]
\caption{The circuit for standart ohmic measurements. The dashed square
represents the sample chamber.}
\label{Fig.1}
\end{figure}
\begin{figure}[tbp]
\caption{The dimensionless $T(\eta )$ dependence given by Eq.(\ref
{Distribution2}) for fixed current,$\lambda $=0;2;5;10 and contact
temperature difference $\Delta T$ found at $\lambda$=0.}
\label{Fig.2}
\end{figure}

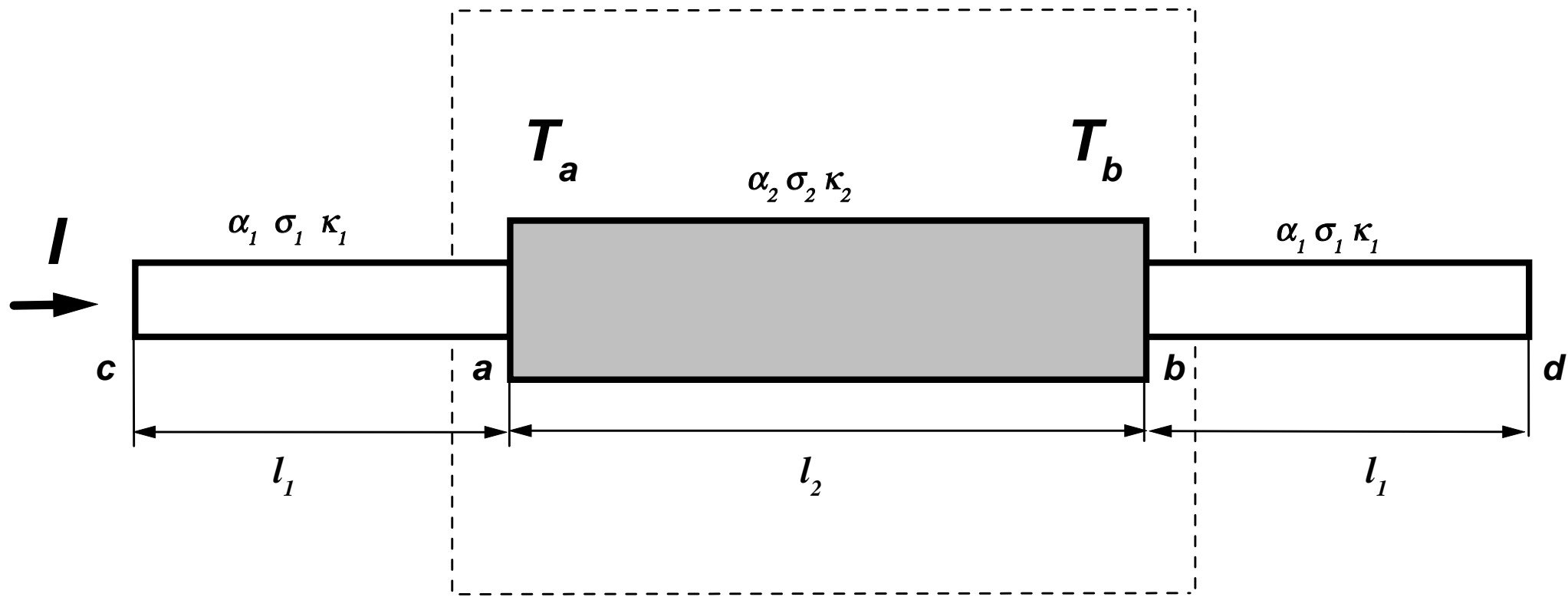
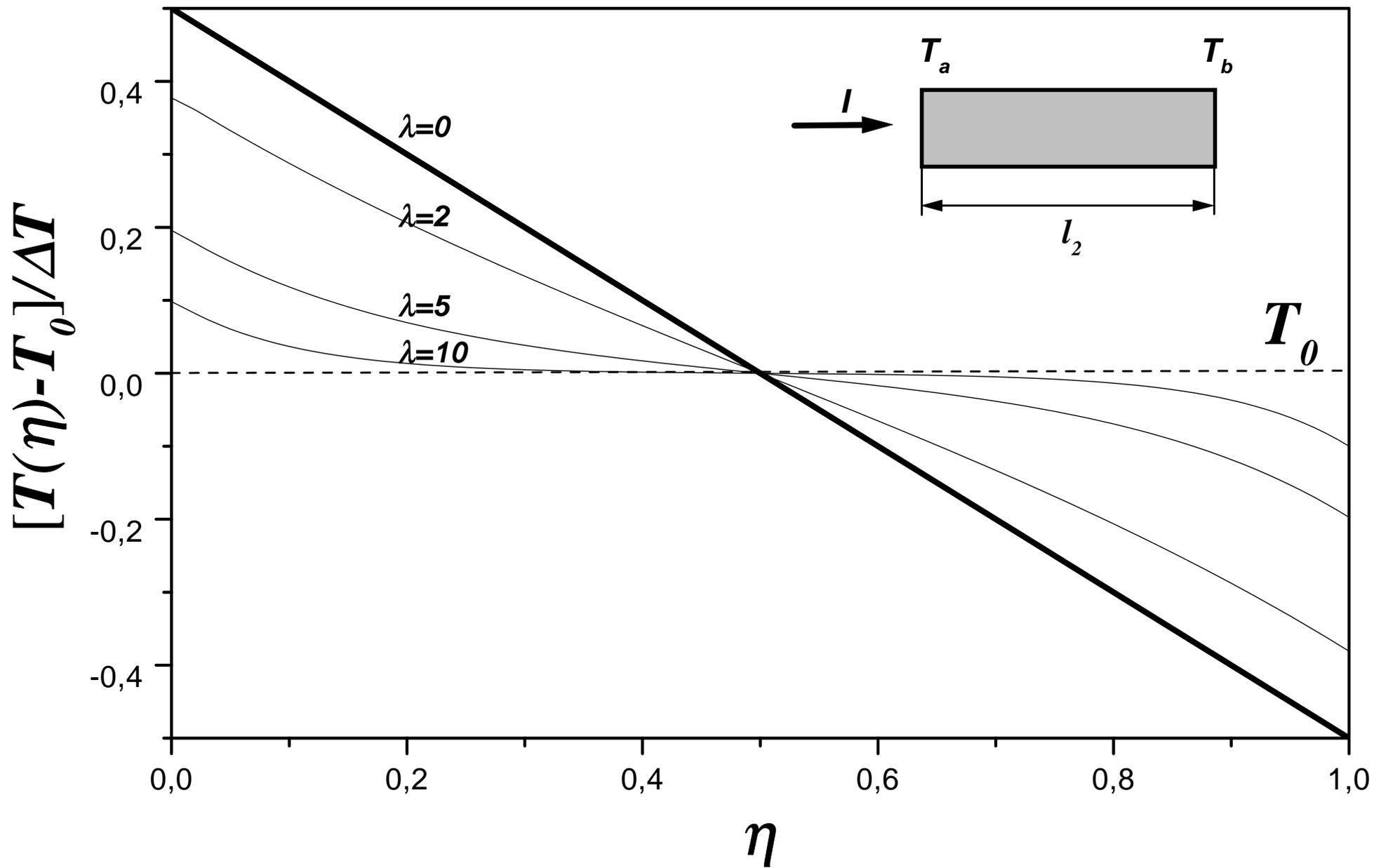
\end{document}